\documentclass[%
 reprint,
 superscriptaddress,
 amsmath,amssymb,
 aps,
 pra,
 longbibliography,
lengthcheck,%
]{revtex4-1}

\usepackage{graphicx}
\usepackage{dcolumn}
\usepackage{bm}
\usepackage{color}
\usepackage{ulem}
\usepackage{hyperref}
\usepackage{braket}

\usepackage{booktabs}
\usepackage{multirow}
\usepackage{amssymb}

\usepackage{diagbox}
\usepackage{needspace}

\begin{document}

\preprint{APS/123-QED}

\title{
	Electrical control of spin photocurrent in a magnetoelectric oxide Cr$_2$O$_3$
}

\author{Zhuo-Cheng Gu}
\affiliation{
	Department of Physics, Institute of Science Tokyo, Meguro, Tokyo, 152-8551, Japan
}

\author{Hiroaki Ishizuka}
\email{ishizuka@hiroishizuka.com}
\affiliation{
	Department of Physics, Institute of Science Tokyo, Meguro, Tokyo, 152-8551, Japan
}

\date{\today}

\begin{abstract}
	Controlling magnetism by electric field or current is a central topic in spintronics.
	In this work, we argue that the magnon spin photocurrent can also be controlled by the electric field in magnetoelectrics. Taking Cr$_2$O$_3$ as an example, we demonstrate how the spin current is modified by the electric field, using nonlinear response theory.
	We find that the Dzyaloshinsky--Moriya interaction induced by the applied field plays a key role in modifying spin-current conductivity, which exhibits pronounced anisotropy with respect to the light polarization.
	In particular, both the resonance frequency and the peak intensity show distinct dependences on the external electric field $E$, demonstrating electrical control of the spin photocurrent. In addition, we show that the two-magnon processes give rise to a continuum spectrum, a consequence of the field-induced spin canting. These results show that Cr$_2$O$_3$ is a promising platform for realizing electrically tunable spin photovoltaic effect.
\end{abstract}

\pacs{
}

\maketitle

\section{Introduction}

Nonlinear optical response offers a powerful avenue for manipulating quantum materials and the quasiparticles they host. Optically controlling magnetism has become a realistic goal in recent years, with important implications for both fundamental physics and practical applications in spintronics~\cite{Baltz2018a,Kirilyuk2018a,Nemec2018a}. Beyond magnetism, spin pumping via magnetoresonance enables optical control of spin currents through the coupling of light to magnetic excitations~\cite{Kajiwara2010a,Heinrich2011a,Ohnuma2014a,Maekawa2017a}. More recently, a qualitatively new mechanism — the magnon spin photocurrent — has been proposed and experimentally observed in collinear antiferromagnets~\cite{Ishizuka2019a,Ishizuka2022a}. This effect, rooted in the Berry phase of magnonic quasiparticles~\cite{Fujiwara2023a,Gu2025a}, generates a dc spin current as a second-order nonlinear response to incident electromagnetic radiation.
The underlying concept has been proven to be general, with the other examples including spinons~\cite{Ishizuka2019b} and phonons~\cite{Ishizuka2024b,Ishizuka2025a}, establishing the generality of the theoretical framework beyond the specific systems considered here.

These phenomena enable interconversion of spin angular momentum and light entirely within insulating systems, avoiding the Joule heating associated with metallic spin-transport devices and opening a route toward ultrafast optical control of spin currents~\cite{Emori2021a}. For practical applications, the ability to tune the resonance frequency and to control the spin current electrically is highly desirable, as it would allow seamless integration with existing electronics. In metallic systems, spin-torque oscillators exploit the spin-transfer torque to control the magnetoresonance frequency electrically~\cite{Boulle2007a,Braganca2010a}; however, analogous functionality in insulating magnets remains largely unexplored. In particular, the electrical tunability of the magnon spin photocurrent has not been addressed. Controlling the spin photocurrent via an applied electric field would provide an all-insulator, electrically reconfigurable optospintronic platform.

In this work, we explore the electrical control of spin photocurrent through the magnetoelectric effect.
Using a primitive four-sublattice spin-chain~\cite{Izuyama1963a} and a three dimension model for Cr$_2$O$_3$~\cite{Samuelsen1970a}, we investigate the nonlinear spin current induced by linearly polarized electromagnetic waves within nonlinear response theory.
We find that the magnon spin photocurrent can be controlled through a Dzyaloshinskii–Moriya interaction (DMI) induced by the applied electric field~\cite{Rado1961a,Hornreich1967a,DeAlcantaraBonfim1980a,Fiebig2005a};
it shifts the resonance frequency of the spin current by one magnon process.
In addition, we show that the two-magnon processes~\cite{Ishizuka2019a} gives rise to a continuum spectrum, a consequence of the field-induced spin canting.
These results demonstrate that the magnon spin photocurrent in magnetoelectric materials can be electrically controlled, offering a promising route to experimentally manipulate it.

This manuscript is organized as follows.
In Sec.~\ref{sec:model}, we introduce the model and method used in this paper.
In Sec.~\ref{sec:1D}, we study a one-dimensional simplified model for Cr$_2$O$_3$,
focusing on how electric-field-induced DMI affects the magnon spin photocurrent.
A three-dimensional model for Cr$_2$O$_3$ is considered in Sec.~\ref{sec:3D}.
Finally, in Sec.~\ref{sec:discussion}, we summarize our findings and discuss the outlook for future research.

\section{Models and Methods}\label{sec:model}
\subsection{Noncollinear magnet}

\begin{figure}[t]
	\includegraphics[width=\linewidth]{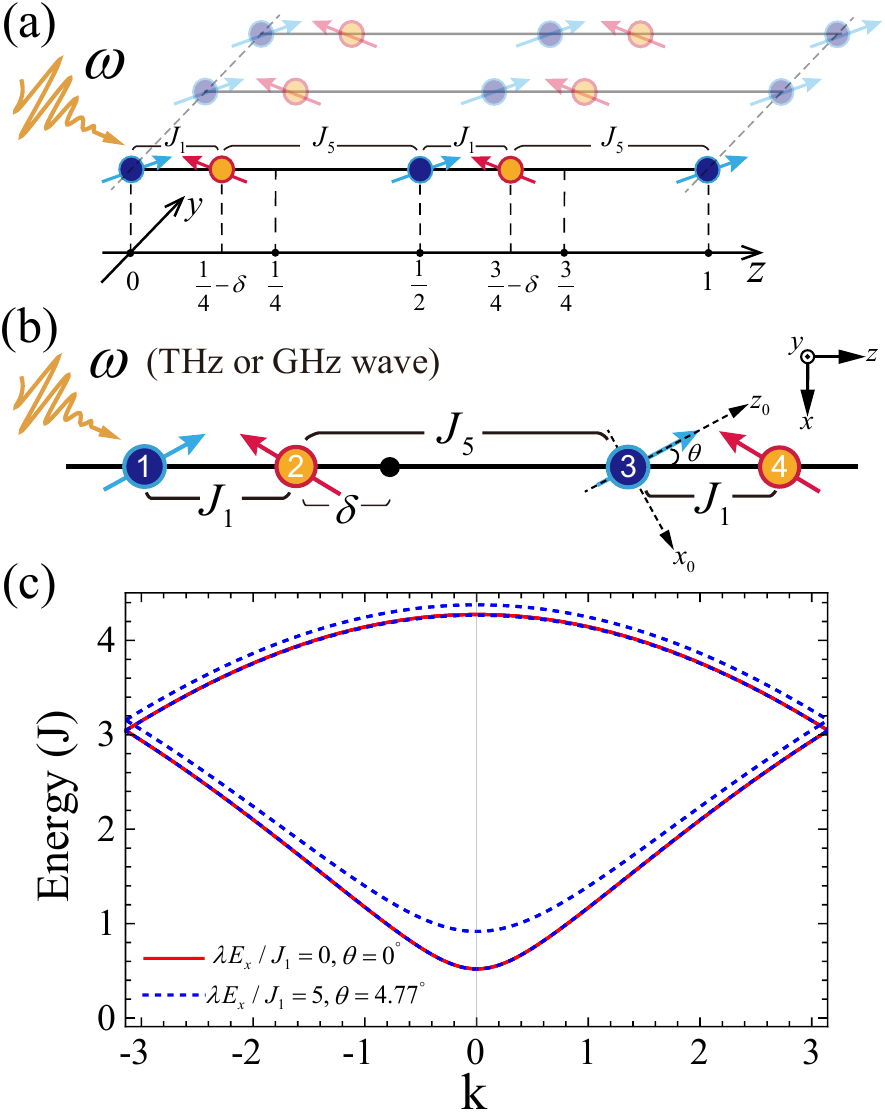}
	\caption{
		A schematic of the spin chain with spin canting and its magnon dispersion. (a) A noncentrosymmetric quasi-one-dimensional model composed of weakly coupled spin chains with the nearest-neighbor interactions, $J_1$ and $J_5$, and a displacement $\delta$. (b) A schematic of the spin chain studied in Sec.~\ref{sec:1D}. The numbers $u=1$--$4$ on each site denote the sublattice, $\theta$ is the angle of canting induced by the DMI, and $(x,y,z)$ and $(x_0, y_0, z_0)$ are the global and local coordinates, respectively.
		(c) Magnon dispersion of the spin chain for $E_x = 0$ and $E_x = 5$ ($\theta = 4.77^{\circ}$) with $\lambda = 0.05$. The results are for $J_1 = -1$, $J_5 = -0.5$, $D_z = 0.005$, and $\delta = 0.1$.
	}\label{fig:model}
\end{figure}

We consider a spin model whose Hamiltonian reads
\begin{align}\label{eq:model_definition_Hamiltonian}
	\mathcal{H} = & - \sum_{\langle jp, lq \rangle} J_{jp,lq}\,\bm S_{jp} \cdot \bm S_{lq} - D_z\sum_{jp} ({S}_{jp}^{z})^2 \nonumber \\
	              & + \sum_{\langle jp, lq \rangle} \bm{\mathcal{D}}_{pq}\cdot\bm{S}_{jp} \times \bm{S}_{lq}.
\end{align}
where $D_z>0$ indicates the uniaxial Ising anisotropy, $J_{jp,lq}$ is the Heisenberg exchange interaction between the two spins, $\bm{S}_{jp}$ and $\bm{S}_{lq}$. Here, $\bm{S}_{ju}=(S_{ju}^{x}, S_{ju}^{y}, S_{ju}^{z})$ is the spin on the sublattice $u$ in the $j$th unit cell.
The last term represents an electric-field-induced DMI, which arises in noncentrosymmetric magnetoelectric systems through spin--orbit coupling. We take the Dzyaloshinskii-Moriya (DM) vector as
$\bm{\mathcal{D}} = \lambda(\bm{e}_{p,q} \times \bm{E})$,
where $\bm e_{pq}$ is the bond-direction unit vector, $E$ is an external electric field, and $\lambda$ denotes the magnetoelectric coupling strength. This term induces a spin canting as shown in Fig.~\ref{fig:model} for a quasi-one-dimensional spin model.

In addition, we consider the Zeeman coupling between the spins and an ac magnetic field, which represents the interaction between light and the spins. The Hamiltonian reads
\begin{align}\label{eq:model_defination_perturbation_operator}
	\mathcal{H}'(t) = - B_\alpha(t) \mu_B g \sum_{j,p} S_{j,p}^\alpha,
\end{align}
where $t$ is the time, $B_\alpha(t)$ ($\alpha =x,y,z$) is the ac magnetic field along the $\alpha$ axis, $g \approx 2$ is the Land\'e g factor, and $\mu_B$ is the Bohr magneton.

The third term in Eq.~\eqref{eq:model_definition_Hamiltonian} arises from the static electric field $\mathbf{E}$, which we use to control the magnetic state and the spin current.
The electric field $\mathbf{E}\parallel \hat{x}$ induces the DMI on each nearest-neighbor bond, whose DM vector reads~\cite{Hornreich1967a,DeAlcantaraBonfim1980a,Fiebig2005a}
\begin{align}
	\bm{\mathcal{D}}_{pq}
	= \lambda(\mathbf{e}_{p,q} \times \mathbf{E})
	= -|\bm{\mathcal{D}}|\,\hat{y}.
\end{align}
The collinear antiferromagnetic order is unstable to the minimal DMI, and hence, with the DMI, the ground state becomes a canted antiferromagnetic order with a small canting angle $\theta$ as illustrated in Fig.~\ref{fig:model}(b).
Here, we define a positive canting angle $\theta$ as a right-handed rotation about the
$\hat y$ axis. For a spin initially polarized along $+\hat z$, this corresponds
to a clockwise rotation in the $xz$ plane when viewed along $-\hat y \to +\hat y$.
In this convention, the chirality vector satisfies $\hat n = +\hat y$, so that $\bm{\mathcal D}\cdot \hat n = -|\bm{\mathcal D}|$.

For the AFM case with only the nearest neighbor interaction $ \sum_{\langle jp,lq \rangle}J_{jp,lq}<0$ (see Fig.~\ref{fig:model}(b)), the canting angle $\theta$ is determined numerically by minimizing the energy,
\begin{align}
	\mathcal{E} = - \sum_{\langle jp,lq \rangle} \left[J_{jp,lq} S^2 \cos\zeta - \left(\bm{\mathcal D_{p,q}} \cdot \hat{n}\right) S^2 \sin\zeta \right]- D_z S^2,
\end{align}
where $\zeta = \pi - 2\theta$. The minimum of this energy functional is given by $\partial\mathcal{E}/\partial\zeta = 0$, from which we find,
\begin{align}\label{eq:models_theta}
	\theta = -\frac{\sum_{\langle jp,lq \rangle}\bm{\mathcal D}_{pq} \cdot \hat{n}}{2\sum_{\langle jp,lq \rangle} J_{jp,lq}}.
\end{align}
Here, the energy minimization gives
$\theta<0$ for sublattices $u=1,3$ and $\theta>0$ for $u=2,4$.
Consequently, the DMI favors an anticlockwise canting of a spin up initially aligned along $+\hat z$.

\subsection{Spin-wave theory}

In the following, we apply the spin-wave theory to the Hamiltonian in Eq.~\eqref{eq:model_definition_Hamiltonian} to study the low-energy dynamics in the presence of an electric field.
To derive the magnon Hamiltonian, we introduce a local coordinate for each spin $(x_0, y_0, z_0)$
(see Fig.~\ref{fig:model}(b)), which are rotated about the $y$ axis by $\theta$ ($-\theta$) for the up (down) sites. The rotation matrix for the clockwise rotation is
\begin{align}\label{eq:method_rotation_transformation}
	 & \begin{pmatrix}
		   S_j^{x} \\
		   S_j^{y} \\
		   S_j^{z}
	   \end{pmatrix}
	=  \begin{pmatrix}
		   \cos\theta  & 0 & \sin\theta \\
		   0           & 1 & 0          \\
		   -\sin\theta & 0 & \cos\theta
	   \end{pmatrix}
	\begin{pmatrix}
		S_j^{x_0} \\
		S_j^{y_0} \\
		S_j^{z_0}
	\end{pmatrix}.
\end{align}
To the linear order in the Holstein-Primakoff transformation, the spin operators are replaced by
\begin{equation}
	\begin{array}{l}
		S_{jp}^{+}  \simeq \sqrt{2S}\,a_{jp},          \,
		S_{jp}^{-}  \simeq \sqrt{2S}\,a_{jp}^{\dagger}, \,
		S_{jp}^{z}  = S - a_{jp}^{\dagger}a_{jp},
	\end{array}
\end{equation}
for the sublattices $p = 1,3$ and
\begin{equation}
	\begin{array}{l}
		S_{jp}^{+}  \simeq \sqrt{2S}\,a_{jp}^{\dagger}, \,
		S_{jp}^{-}  \simeq \sqrt{2S}\,a_{jp},           \,
		S_{jp}^{z}  = a_{jp}^{\dagger}a_{jp}-S,
	\end{array}
\end{equation}
for $p = 2,4$, where $S$ is a spin of $S_j$. Here, $a_{jp}$ ($a_{jp}^{\dagger}$) is the magnon annihilation (creation) operator in the $j$th unit cell.

The resultant spin-wave Hamiltonian reduces to a free boson Hamiltonian of the Bogoliubov-de Gennes (BdG) form,
\begin{align}
	\mathcal{H} = \frac12 \sum_{\bm k} \phi_{\bm k}^\dagger \hat{H}_{\text{BdG}}(\bm k) \phi_{\bm k},
\end{align}
and $\hat{H}_{\text{BdG}}(\bm k)$ reads,
\begin{align}
	\hat{H}_{\text{BdG}}(\bm k) \equiv
	\begin{pmatrix}
		\Xi_{\bm k}            & \Delta_{\bm k} \\
		\Delta_{\bm k}^\dagger & \Xi^T_{-\bm k}
	\end{pmatrix}
	= \begin{pmatrix}
		  \Xi_{\bm k}    & \Delta_{\bm k} \\
		  \Delta_{\bm k} & \Xi_{\bm k}
	  \end{pmatrix},
\end{align}
where we assume $\Delta_{\bm k}^\dagger = \Delta_{\bm k}$ and $ \Xi^T_{-\bm k} = \Xi_{\bm k}$. Here, $\phi_{\bm k}=\left(a_{1,\bm k},\dots,a_{n_{uc},\bm k},a_{1,-\bm k}^\dagger,\dots,a_{n_{uc},-\bm k}^\dagger\right)^T$ is an array of boson annihilation (creation) operators $a_{j\bm k}$ ($a_{j\bm k}^\dagger$) with $\bm k$ being the momentum and $j$ denoting the sublattice. Here, $a_{\bm k} \equiv (1/\sqrt{N_c})\sum_{r} e^{i\bm k \cdot \bm r} a(\bm r)$ is the Fourier transformation of Holstein-Primakoff bosons, where $N_c$ is the number of unit cells.
This Hamiltonian is readily diagonalized by a well-known protocol~\cite{Colpa1978a},
\begin{align}
	\Sigma^z \hat{H}_{\text{BdG}}(\bm k) u_{\pm}(\bm k) = \omega_{\pm}(k) u_{\pm}(\bm k),
\end{align}
where $\Sigma^z = \sigma^z \otimes \mathcal{I}$ and $\sigma^z$ is the Pauli matrix. Finally, the Hamiltonian becomes
\begin{align}
	\mathcal{H} = \frac12 \sum_{\bm k} \psi_{\bm k}^\dagger \hat{E}_{\bm k} \psi_{\bm k},
\end{align}
where $\hat{E}_{\bm k}$ is a diagonalized matrix and $\psi_{\bm k} = (\alpha_{j,\bm k}, \alpha_{j, -\bm k}^\dagger)^T$. In what follows, we mainly discuss the spin current within the linear spin wave theory.

\subsection{Spin Photocurrent}

In this work, we define the spin current operator for $S^z$ following the standard convention,
\begin{align}\label{eq:model_defination_spin_current_operator}
	\mathcal{J}^z = \frac{1}{4N_c} \sum_{\langle jp,lq \rangle} J_{jp,lq}\left(\bm{r}_{jp} - \bm{r}_{lq}\right)_z \left(S_{jp}^{x} S_{lq}^{y} - S_{jp}^{y} S_{lq}^{x} \right).
\end{align}
This is the equation for spin current derived from the continuity equation of spin angular momentum, when it is conserved.
Strictly speaking, the DMI breaks the angular momentum conservation through spin canting.
However, as the induced canting is small in our case, Eq.~\eqref{eq:model_defination_spin_current_operator} is expected to be a good approximation to the leading order in $E$.

In this paper, we focus on the second-order nonlinear conductivity for the spin current, $
	[\sigma^{(2)}]_{\nu\lambda}^\mu$, which is the lowest-order contribution to the dc spin photocurrent induced by the electromagnetic field. Here, $[\sigma^{(2)}]_{\nu\lambda}^\mu$ describes the generation of a spin photocurrent flowing along the $\mu$ direction in response to an incident electromagnetic field with polarization components along the $\nu$ and $\lambda$ directions, which is defined by
\begin{align}\label{eq:general_spincurrent_defination}
	J_\mu (\Omega) = \int \sum_{\nu \lambda} \sigma^{(2)}_{\mu \nu \lambda} (\Omega;\omega,-\omega) B_{\nu}(\omega) B_{\lambda}(\Omega-\omega) d\omega,
\end{align}
where $J_\mu(\Omega) = \int \langle \mathcal J_{\mu} \rangle e^{-i\Omega t}dt$ with $\langle O\rangle$ being the thermal average of $O$, and $B_\alpha(\omega) = \int B_\alpha(t) e^{-i\omega t	}dt$. For the external field in Eq.~\eqref{eq:model_defination_perturbation_operator} and~\eqref{eq:model_perturbation_final}, the nonlinear conductivity reads
\begin{align}
	\sigma_{\mu\nu\lambda}^{(2)}(0;\omega,-\omega) = \left[\sigma_{\mu\nu\lambda}^{(2)}\right]_{\mathrm{one}} + \left[\sigma_{\mu\nu\lambda}^{(2)}\right]_{\mathrm{two}}.
\end{align}
Here, $[\sigma_{\mu\nu\lambda}^{(2)}]_{\mathrm{one}}$ is the contribution of the one-magnon process~\cite{Ishizuka2022a},
\begin{widetext}
	\begin{align}\label{eq:model_definition_spin_conductivity_one}
		 & \left[\sigma_{\mu\nu\lambda}^{(2)}\right]_{\mathrm{one}} = -\frac{1}{4\pi V} \sum_{l = 1}^{2N} \sum_{m = 1}^{2N}\frac{\gamma_l \gamma_m}{\hbar\omega-\gamma_l E_{l \bm 0}-i\eta} \frac{[\tilde{\beta}_{\bm 0}^\nu]_l [\tilde{\beta}_{\bm 0}^\lambda]_{m+N\gamma_m} \left([\tilde{\mathcal{J}}_{\bm 0}^\mu]_{l,m}+[\tilde{\mathcal{J}}_{\bm 0}^\mu]_{m+N\gamma_m,l+N\gamma_l}\right)}{-\gamma_l E_{l \bm 0} +\gamma_m E_{m \bm 0} -i\eta},
	\end{align}
	and $[\sigma_{\mu\nu\lambda}^{(2)}]_{\mathrm{two}}$ is the two-magnon process~\cite{Ishizuka2019a} (see Appendix~\ref{sec:appendix_derive_conductivity_TMP} for details),
	\begin{align}\label{eq:model_definition_spin_conductivity_two}
		 & \left[\sigma_{\mu\nu\lambda}^{(2)}\right]_{\mathrm{two}} = \frac{1}{32\pi V} \sum_{\bm k} \sum_{n,m,l=1}^{2N}
		\frac{\gamma_l(\gamma_m  - \gamma_n)}
		{\hbar\omega + \gamma_n E_{n \bm k} - \gamma_m E_{m \bm k} - i\eta} \nonumber                                    \\
		 &
		\times \left[
		\frac{[\tilde{\beta}_{\bm k}^\nu]_{nm} [\tilde{\beta}_{\bm k}^\lambda]_{n+\gamma_n N, l} \left([{\tilde{\mathcal{J}}}_{\bm k}^\mu]_{m,l+\gamma_l N} + [{\tilde{\mathcal{J}}}_{\bm k}^\mu]_{l,m+\gamma_m N} \right)}
		{ - \gamma_l E_{l\bm k} - \gamma_m E_{m\bm k} - i\eta}
		-\frac{[\tilde{\beta}_{\bm k}^\nu]_{nm} [{\tilde{\beta}}_{\bm k}^\lambda]_{l n} \left( [\tilde{\mathcal{J}}_{\bm k}^\mu]_{m l} + [\tilde{\mathcal{J}}_{\bm k}^\mu]_{l+\gamma_l N, m+\gamma_m N} \right)}
		{\gamma_l E_{l\bm k} - \gamma_m E_{m\bm k} - i\eta} \right.
		\nonumber                                                                                                        \\
		 & \left.\quad
		- \frac{[\tilde{\beta}_{\bm k}^\nu]_{nm} {[\tilde{\mathcal{J}}_{\bm k}^{\mu}]}_{n+\gamma_n N, l} \left([\tilde{\beta}_{\bm k}^\lambda]_{m,l+\gamma_l N} + [\tilde{\beta}_{\bm k}^\lambda]_{l,m+\gamma_m N} \right)}
		{\gamma_l E_{l\bm k} + \gamma_n E_{n\bm k} - i\eta}
		+ \frac{[\tilde{\beta}_{\bm k}^\nu]_{nm} [\tilde{\mathcal{J}}_{\bm k}^\mu]_{l n} \left([\tilde{\beta}_{\bm k}^{\lambda}]_{m l} +[\tilde{\beta}_{\bm k}^\lambda]_{l+\gamma_l N, m+\gamma_m N} \right)}
		{\gamma_n E_{n\bm k} - \gamma_l E_{l\bm k} - i\eta}
		\right],
	\end{align}
\end{widetext}
where $\hbar$ is the reduced Planck constant, $V$ is the volume of the system, $N$ is the number of sublattices, and
\begin{align}\label{eq:method_gamma}
	\gamma_l= & \left\{\begin{array}{rl}
		                   1  & (1 \leq l \leq N)     \\
		                   -1 & (N+1 \leq l \leq 2N).
	                   \end{array}\right.
\end{align}
In the braces, $[\tilde{\mathcal{J}}_{\bm k}^\mu]_{lm} = \langle l |\tilde{\mathcal{J}}_{\bm k}^\mu| m \rangle$ is the matrix elements for $\tilde{\mathcal{J}}_{\bm k}^\mu$ where $|m \rangle$ is the ket vector of $m$th magnon state with $\bm k$, $E_{m \bm k}$ is the eigenenergy of $|m \rangle$, and $\eta$ is the inverse magnon relaxation time. Matrix elements $[\tilde \beta_{\bm k}^\nu]_{nm}$ and $[\tilde \beta_{\bm 0}^\nu]_{l}$ are the coefficient of $H'$ in the eigenstate basis defined by
\begin{align}\label{eq:model_perturbation_final}
	 & \mathcal{H}'(t)  =  - \sum_{\substack{l                                                                  \\\nu=x_0,y_0,z_0}}^N B_{\nu}(t)\left([\tilde{\beta}_{\bm 0}^{\nu}]_{l}\alpha_{\bm 0l}+[\tilde{\beta}_{\bm 0}^{\nu}]_{l+N} \alpha_{\bm 0l}^\dagger\right) \nonumber \\
	 & - \sum_{\substack{n,m,\bm k                                                                              \\\nu=x_0,z_0}}^N  B_{\nu}(t) \left([\tilde{\beta}_{\bm k}^{\nu}]_{nm} \alpha_{\bm k n}^\dagger \alpha_{\bm k m}
	+[\tilde{\beta}_{\bm  k}^{\nu}]_{n,m+N} \alpha_{\bm k n}^\dagger \alpha_{\bm k m}^\dagger \right. \nonumber \\
	 & \left.
	+[\tilde{\beta}_{\bm  k}^{\nu}]_{n+N,m} \alpha_{\bm k n} \alpha_{\bm k m}
	+[\tilde{\beta}_{\bm  k}^{\nu}]_{n+N,m+N} \alpha_{\bm k n} \alpha_{\bm k m}^\dagger\right),
\end{align}
with $[\tilde{\beta}_{\bm 0}^\nu]_{n+N} = ([\tilde{\beta}_{\bm 0}^\nu]_{n})^*$ and $[\tilde{\beta}_{\bm k}^\nu]_{n+N,m+N} = ([\tilde{\beta}_{\bm k}^\nu]_{nm})^*$.

\section{One-dimensional spin chain}\label{sec:1D}
First, we study a spin chain, which was introduced as an effective model for Cr$_2$O$_3$~\cite{Izuyama1963a}; a schematic of the spin chain is shown in Fig.~\ref{fig:model}(b).
The Hamiltonian reads,
\begin{align}
	\label{eq:1D_Hamiltonian_initial}
	 & \mathcal{H}^{1D} = - \sum_{\langle ip, jq \rangle} \left(\mathcal{H}_{NN}^{1D} - \mathcal{H}_{\text{DMI}}^{1D}\right) + \mathcal{H}_D, \nonumber \\
	 & \mathcal{H}_{NN}^{1D} = {J_1}{{\bm S}_{j,1}} \cdot {{\bm S}_{j,2}} + {J_5}{{\bm S}_{j,2}} \cdot {{\bm S}_{j,3}} \nonumber                        \\
	 & \qquad \qquad + {J_1}{{\bm S}_{j,3}} \cdot {{\bm S}_{j,4}} + {J_5}{{\bm S}_{j,4}} \cdot {{\bm S}_{j + 1,1}}, \nonumber                           \\
	 & \mathcal{H}_{\text{DMI}}^{1D} = \bm{\mathcal{D}} \cdot \left(\bm{S}_{j,1} \times \bm{S}_{j,2} + \bm{S}_{j,3} \times \bm{S}_{j,4}\right),         \\
	 & \mathcal{H}_D = - {D_z}{\sum\limits_{j,p} {(S_{j,p}^z)} ^2},
\end{align}
where $\mathcal{H}_{NN}^{1D}$ is the nearest neighbor (NN) interaction, $\mathcal{H}_{\text{DMI}}^{1D}$ is the DMI between the spins and the external electromagnetic field, and $\mathcal{H}_D$ denotes the uniform anisotropy. In the following, we assume the nearest-neighbor interactions to be $J_1,\;J_5 < 0$.

\begin{figure*}[t]
	\includegraphics[width=\linewidth]{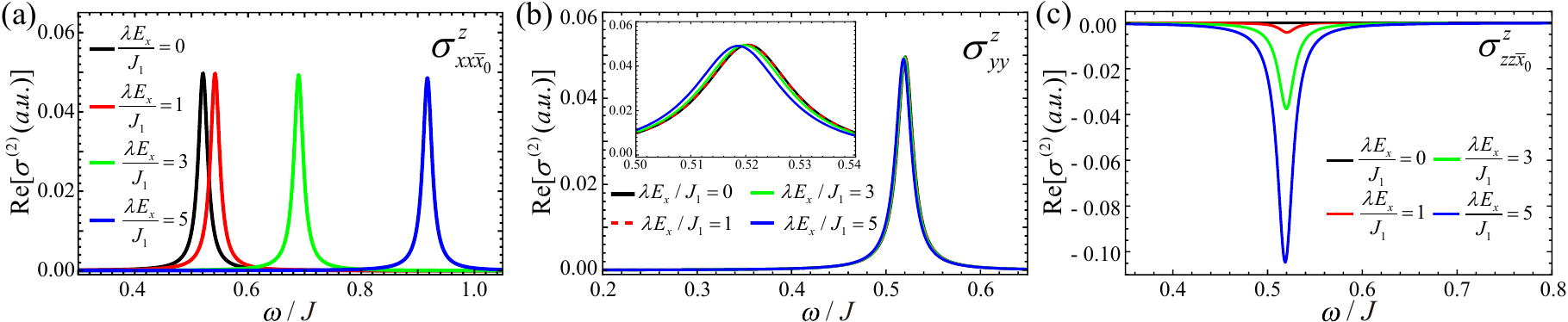}
	\caption{
	Nonlinear spin current conductivity for $S^z$ through the one-magnon process along the $\alpha$ ($\alpha = x_0,y_0,z_0$) direction under a static electric field $E_x$. Frequency dependence of (a) $[\sigma^{(2)}]^z_{xx \bar x_0}$, (b) $[\sigma^{(2)}]^z_{yy}$, and (c) $[\sigma^{(2)}]^z_{zz \bar x_0}$. The results are obtained for $J_1 = -1$, $J_5 = -0.5$, $D_z = 0.005$, $\delta = 0.1$, $\eta = \omega_1(k=0, E=0) \approx 0.01$, and $\lambda = 0.05$. The canting angles corresponding to $\lambda E_x/J_1 = 1$, 3, and 5 are $\theta = 0.95^\circ$, $2.86^\circ$, and $4.77^\circ$, respectively.
	}\label{fig:1D_conductivity_oneprocess}
\end{figure*}

Within the linear spin-wave approximation, the magnon Hamiltonian for $H^{1D}$ becomes
\begin{widetext}
	\begin{align}\label{eq:1D_Hamiltonian_final_form}
		\mathcal{H} = & S \sum_k \left[D_z \left(3\cos^2\theta - 1\right) + \lambda E_x \sin2\theta - \left(J_1 + J_5\right)\cos2\theta \right]\left(a_k^\dagger a_k + b_{-k}^\dagger b_{-k} + c_k^\dagger c_k + d_{-k}^\dagger d_{-k} \right) \nonumber                                                                                                                   \\
		              & -\frac12 S \sum_k \left\{\left[J_1 \left(\cos2\theta+1\right) - \lambda E_x \sin2\theta \right] e^{-ik\left(\frac14-\delta\right)} \left(a_k b_{-k} + c_k d_{-k} \right) + J_5\left(\cos2\theta+1\right) e^{ik\left(\frac14 + \delta\right)}\left(c_k b_{-k} + a_k d_{-k}\right) \right\}\nonumber                                                 \\
		              & -\frac12 S \sum_k \left[J_1 \left(\cos2\theta+1\right) - \lambda E_x \sin2\theta \right] e^{ik\left(\frac14-\delta\right)}\left(a_k^\dagger b_{-k}^\dagger + c_k^\dagger d_{-k}^\dagger \right) + J_5\left(\cos2\theta+1\right) e^{-ik\left(\frac14 + \delta\right)}\left(c_k^\dagger b_{-k}^\dagger + a_k^\dagger d_{-k}^\dagger \right)\nonumber \\
		              & -\frac12 S \sum_k \left[J_1 \left(\cos2\theta-1\right) - \lambda E_x \sin2\theta\right] e^{-ik\left(\frac14-\delta\right)} \left(a_k b_{k}^\dagger + c_k d_k^\dagger\right) + J_5 \left(\cos2\theta - 1 \right) e^{ik\left(\frac14 + \delta\right)}\left(c_k b_k^\dagger + a_k d_k^\dagger \right)
		\nonumber                                                                                                                                                                                                                                                                                                                                                          \\
		              & -\frac12 S \sum_k \left[J_1 \left(\cos2\theta-1\right) - \lambda E_x \sin2\theta \right]e^{ik\left(\frac14-\delta\right)} \left(a_k^\dagger b_k + c_k^\dagger d_k \right) + J_5 \left(\cos2\theta - 1 \right) e^{-ik\left(\frac14 + \delta\right)}\left(c_k^\dagger b_k + a_k^\dagger d_k \right) \nonumber                                        \\
		              & - \frac12 D_z S\sum_k \left(a_k^\dagger a_{-k}^\dagger + a_k a_{-k} + b_k^\dagger b_{-k}^\dagger + b_k b_{-k} + c_k^\dagger c_{-k}^\dagger + c_k c_{-k} + d_{k}^\dagger d_{-k}^\dagger + d_{k} d_{-k}\right)\sin^2\theta,
	\end{align}
	where we have neglected the constant terms.
	Here, $a_{k}$ ($a_{k}^\dagger$), $b_{k}$ ($b_{k}^\dagger$), $c_{k}$ ($c_{k}^\dagger$), and $d_{k}$ ($d_{k}^\dagger$) are the annihilation (creation) operators for the magnons on sublattices 1--4, respectively.
	Similarly, the spin current operator reads
	\begin{align}\label{eq:1D_spincurrent_final_form}
		\mathcal{J}^z & = \frac{iS J_1\cos\theta}{2N}\left(\frac14 - \delta\right) \sum_{k}\left[\left(e^{ik\left(\frac14 - \delta\right)}b_{-k}^\dagger a_k^\dagger - e^{-ik\left(\frac14 - \delta\right)}b_{-k} a_k \right) + \left(e^{ik\left(\frac14 - \delta\right)}d_{-k}^\dagger c_k^\dagger - e^{-ik\left(\frac14 - \delta\right)}d_{-k} c_k \right) \right] \nonumber \\
		              & + \frac{iS J_5\cos\theta}{2N}\left(\frac14 + \delta\right) \sum_k \left[\left(e^{ik\left(\frac14 + \delta\right)} c_k b_{-k} - e^{-ik\left(\frac14 + \delta\right)} c_k^\dagger b_{-k}^\dagger \right) + \left(e^{ik\left(\frac14 + \delta\right)} a_k d_{-k} - e^{-ik\left(\frac14 + \delta\right)} a_k^\dagger d_{-k}^\dagger \right) \right].
	\end{align}
	\needspace{12\baselineskip}
\end{widetext}

The perturbation operator in Eq.~\eqref{eq:model_defination_perturbation_operator} reads,
\begin{subequations}\label{eq:model_perturbation_total}
	\begin{align}
		 & H_x'(t) = - B_x(t) \nonumber                                                                                                                                                                                             \\
		 & \times \sum_{p,\bm k} \left[\frac{\sqrt{2SN}}{2}\cos\theta (p_{\bm k} + p_{\bm k}^\dagger)\delta_{\bm 0, \bm k} + \sin\theta \, p_{\bm k}^\dagger p_{\bm k} \right],
		\label{eq:model_perturbation_x}                                                                                                                                                                                             \\
		 & H_y'(t) = -\frac{\sqrt{2SN}}{2i} B_y(t) \sum_{p, \bm k} \xi_p (p_{\bm k} - p_{\bm k}^\dagger) \delta_{\bm 0, \bm k}, \label{eq:model_perturbation_y}                                                                     \\
		 & H_z'(t) = - B_z(t) \nonumber                                                                                                                                                                                             \\
		 & \times \sum_{p,\bm k} \xi_p \left[\frac{\sqrt{2SN}}{2} \sin\theta  \left(p_{\bm k} + p_{\bm k}^\dagger \right) \delta_{\bm 0, \bm k} + \cos\theta \, p_{\bm k}^\dagger p_{\bm k}\right], \label{eq:model_perturbation_z}
	\end{align}
\end{subequations}
where $p=a,b,c,d$ labels sublattices and $\xi_p = \pm1$ encodes sublattice-dependent signs. The terms containing $\delta_{\bm0,\bm k}$ describe one-magnon process. In $H_x'$ and $H_z'$, the first and second terms correspond to the projection of the light field onto the local $x_0$- and $z_0$-axis components, respectively. Compared to the previous study of collinear spin configuration~\cite{Gu2025a}, the component along the $z_0$ direction allows non-zero two-magnon process spin photocurrent~\cite{Ishizuka2019a} to be generated, in addition to the one-magnon process ones~\cite{Ishizuka2022a,Gu2025a}. After performing the Bogoliubov transformation, $H'$ in the eigenstate basis is obtained as Eq.~\eqref{eq:model_perturbation_final}.

We show the dispersion of the magnon bands in Fig.~\ref{fig:model}(c). There are two degenerate magnon branches in the absence of an external field. With a non-zero electric field $E_x$, the degeneracy is lifted by the DMI and results in four different magnon bands, as shown in Fig.~\ref{fig:model}(c). Notably, the splitting is highly asymmetric; two branches remain nearly unchanged and nearly overlap with the zero-field dispersion, whereas the other two bands shift to a higher energy.
This behavior is distinct from the Zeeman splitting, in which two degenerate modes are typically displaced in opposite directions. The asymmetric reconstruction originates from the anisotropic coupling of the DMI~\cite{Kawano2019a}, whose effect depends sensitively on the relative orientations of the DM vector, the ordered spins, and the magnon propagation direction.

\subsection{One-magnon process}
The frequency dependence of the conductivity for the one-magnon process is shown in Fig.~\ref{fig:1D_conductivity_oneprocess}.
We study the conductivity generating by linearly polarized light along $x_0$ direction in the local frame ($\sigma_{xx \bar x_0}^z$ and $\sigma_{zz\bar x_0}^z$, shown in Figs.~\ref{fig:1D_conductivity_oneprocess}(a) and (c)), and along the $y$ direction in the global frame ($\sigma_{yy}^z$, shown in Fig.~\ref{fig:1D_conductivity_oneprocess}(b)).
Here, the resonance peaks correspond to the magnon frequencies at $k=0$, as can be inferred from Eq.~\eqref{eq:model_definition_spin_conductivity_one}. It shows that a dc spin current is generated through a nonlinear optical process as argued in a recent paper~\cite{Gu2025a}.

As shown in Fig.~\ref{fig:1D_conductivity_oneprocess}, $\sigma^z$ shows a polarization dependence distinct from the collinear case~\cite{Gu2025a}, in which the spectrum of nonlinear conductivity for $\sigma^{z}_{xx}$ and $\sigma^{z}_{yy}$ are the same. In contrast, due to the absence of rotation symmetry about the $z$ axis induced by the electric field, the two conductivities show different $\omega$ dependence.
This single-peak structure can be understood from the selection rules encoded in the perturbation operators after the Bogoliubov transformation, $\tilde{\beta}_{\bm 0}^{\alpha}$ ($\alpha=x,y,z$). Similar to the collinear spin case~\cite{Gu2025a}, only specific matrix elements remain nonzero. Consequently, each conductivity channel couples selectively to a particular magnon branch. Specifically, the operators take the form,
\begin{align}\label{eq:1D_beta_form}
	 & \tilde\beta_{\bm 0}^{x} =
	(0,[\tilde{\beta}_{\bm 0}^{\nu}]_2,0,0,0,[\tilde{\beta}_{\bm 0}^{\nu}]_2,0,0), \\
	 & \tilde\beta_{\bm 0}^{y, z} =
	([\tilde{\beta}_{\bm 0}^{\nu}]_1,0,0,0,[\tilde{\beta}_{\bm 0}^{\nu}]_1,0,0,0).
\end{align}
From these expressions, we see that the $x$-polarized channel with only nonzero matrix elements associated with the second magnon branch survives, while the $y$- and $z$-polarized channels are governed by the first branch. This selective coupling yields the resonance peaks at different frequencies in $\sigma_{xx\bar x_0}^z$, $\sigma_{yy}^z$, and $\sigma_{zz\bar x_0}^z$, which are distinct from the case with a magnetic field where two resonance peaks appear~\cite{Gu2025a}.

Importantly, the selection rules is a consequence of the sublattice-dependent structure of $H_y'$ and $H_z'$ in Eq.~\eqref{eq:model_perturbation_total}.
In particular, the $y$- and $z$-polarized perturbations contain the sublattice factor $\xi_p$, whereas the $x$-polarized perturbation does not. As a result, the $x$-polarized perturbation couples to a different magnon band from the $y$- and $z$-polarized perturbations, leading to different overlaps with the magnon eigenmodes.

The branch-selective optical coupling explains the difference in the electric-field effect.
As shown in Fig.~\ref{fig:model}(c), the $E_{0\bm k=\bm0}$ (which is excited by $B_y$ and $B_z$) changes only slightly under the applied electric field and can therefore be regarded as nearly constant, whereas $E_{1\bm k=\bm0}$ shifts to a higher energy.
As a result, for weak electric fields, such as $E_x=1$,
the peak position of $\sigma_{xx \bar x_0}^z$ shifts by applying the electric field while those for $\sigma_{yy \bar x_0}^z$ and $\sigma_{zz \bar x_0}^z$ remains the same.

\subsection{Two-magnon process}
\begin{figure}
	\includegraphics[width=\linewidth]{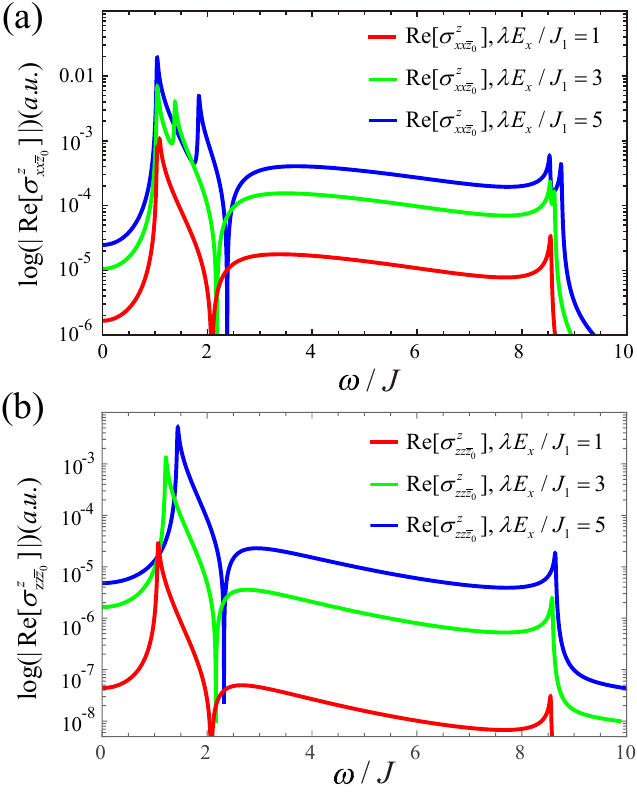}
	\caption{
		Nonlinear spin current conductivity for the 1D model in Eq.~\ref{eq:1D_Hamiltonian_initial} induced by the two-magnon process; (a) $\sigma_{xx \bar z_0}^{z}$ and (b) $\sigma_{zz \bar z_0}^{z}$. Each line is the conductivity with different electric fields $E =1$(red), 3 (green), and 5 (blue). The canting angles corresponding to $\lambda E_x/J_1 = 1$, 3, and 5 are $\theta = 0.95^\circ$, $2.86^\circ$, and $4.77^\circ$, respectively. Frequency dependence of the spin conductivity for $J_1 = -1$, $J_5 = -0.5$, $D_z = 0.005$, $\delta = 0.1$, $\lambda = 0.05$, and $\eta = \omega_1(k=0, E=0) \approx 0.01$.
	}\label{fig:1D_conductivity_twoprocess}
\end{figure}

Figure~\ref{fig:1D_conductivity_twoprocess} shows the frequency dependence of the nonlinear conductivities for the two-magnon process, $\sigma_{xx\bar z_0}^z$ and $\sigma_{zz\bar z_0}^z$. These conductivities are generated by the coupling of linearly polarized light (along the $x$ and $z$ directions, respectively) to the $z_0$ component. In contrast to the sharp peaks typical of the one-magnon process, the two-magnon process shows a continuum, reflecting
the simultaneous creation of two magnons with opposite momenta ($k$ and $-k$). The total energy of the two-magnon excitation is given by $\omega(k_1) + \omega(k_2)$. Due to the requirement of momentum conservation ($k_1 + k_2 \approx 0$), the resonance condition simplifies to $\omega \in [\omega_1(0)+\omega_2(0), \omega_3(0)+\omega_4(0)]$. Consequently, as $k$ probes the entire Brillouin zone, the resulting conductivity spectrum spans a frequency range approximately twice the bandwidth of magnon dispersion.

The nonlinear conductivity for the two-magnon process in one-dimensional systems can be expressed as (Appendix~\ref{sec:appendix_conductivity_DOS}),
\begin{align}
	\mathrm{Re}\left[\sigma_{\mu\nu\lambda}^{(2)}(\omega)\right]
	\propto
	\sum_{nml}\sum_{k_i}
	\frac{
	\mathcal W_{nml}(k_i)
	}{
	\left|
	\partial_k \Delta_{mn}(k)
	\right|_{k=k_i}
	},
	\label{eq:maintext_1d_conductivity_dos}
\end{align}
where $k_i$ satisfies $\Delta_{mn}(k_i)=\hbar\omega$. This form indicates that the conductivity inherits the same singular structure as the one-dimensional joint density of states (JDOS), while the overall spectral weight is further modulated by the matrix-element factor $\mathcal W_{nml}(k_i)$. Therefore, when $\mathcal W_{nml}(k)$ varies slowly around the resonant momenta, the conductivity approximately follows the DOS, whereas noticeable deviations can arise when the matrix elements exhibit strong momentum dependence.

In $\sigma_{xx \bar z_0}^{z}$ [Fig.~\ref{fig:1D_conductivity_twoprocess}(a)], the two-magnon continuum shows two peaks in the $0\le\omega \le2$ window and another pair of peaks in $8\le\omega\le9$. As the electric-field strength increases, these features gradually evolve into a well-resolved multi-peak continuum structure.
This evolution closely tracks the corresponding changes in the density of states. In particular, the prominent spectral features in $\sigma_{xx \bar z_0}^{z}$ show a clear one-to-one correspondence with the van Hove singularities of the JDOS associated with the two-magnon excitation energy.
Such behavior is consistent with Eq.~\eqref{eq:maintext_1d_conductivity_dos}, and is analogous to the DOS-governed structures reported in previous studies of nonlinear spin photocurrent for the two-magnon process~\cite{Ishizuka2019a,Ishizuka2019b}.
Physically, this reflects the DMI-induced modification of the magnon dispersion discussed above [Fig.~\ref{fig:model}(c)]. In particular, the DMI lifts the degeneracy of the magnon bands and splits the dispersion near $k=0$, thereby reshaping it.
In the vicinity of the $k=0$ region, the magnon group velocity becomes small, or equivalently, the bands become locally flat. This results in an enhancement of the JDOS, giving rise to multiple van Hove points, which are reflected as distinct multi-peak continuum structures in the two-magnon optical response.
From an equivalent viewpoint, the magnon bands become locally flat in this momentum region, leading to an enhanced JDOS and hence an enhanced optical response.

In contrast to $\sigma_{xx \bar z_0}^{z}$, $\sigma_{zz \bar z_0}^{z}$ exhibits qualitatively different spectral behavior and does not show a simple one-to-one correspondence with the JDOS associated with $2\omega(k)$ [Fig.~\ref{fig:1D_conductivity_twoprocess}(b)]. This difference originates from the distinct symmetry structure of the perturbation operators in Eqs.~\eqref{eq:model_perturbation_x} and~\eqref{eq:model_perturbation_z}. In particular, the $z$-polarized perturbation $H_z'$ contains the staggered sublattice factor $\xi_p=\pm1$, which leads to destructive interference of the matrix-element factor $\mathcal{W}_{nml}(k_i)$ at the van Hove momenta ($k=0$).
Consequently, the corresponding peak continuum structure vanishes at the JDOS singularities, suppressing that at $2\omega(k=0)$. The spectral weight $\mathcal W_{nml}(k_i)$ is therefore shifted away from the singular points and redistributed into neighboring frequency regions. By contrast, the $x$-polarized channel does not suffer from this staggered cancellation, so its peak and multi-peak continuum structures remain aligned with the van Hove points. In addition, the overall magnitude of $\sigma_{xx\bar z_0}^{z}$ is approximately one order of magnitude larger than that of $\sigma_{zz \bar z_0}^{z}$.

\section{Chromium oxide}\label{sec:3D}
\begin{figure}[h]
	\includegraphics[width=\linewidth]{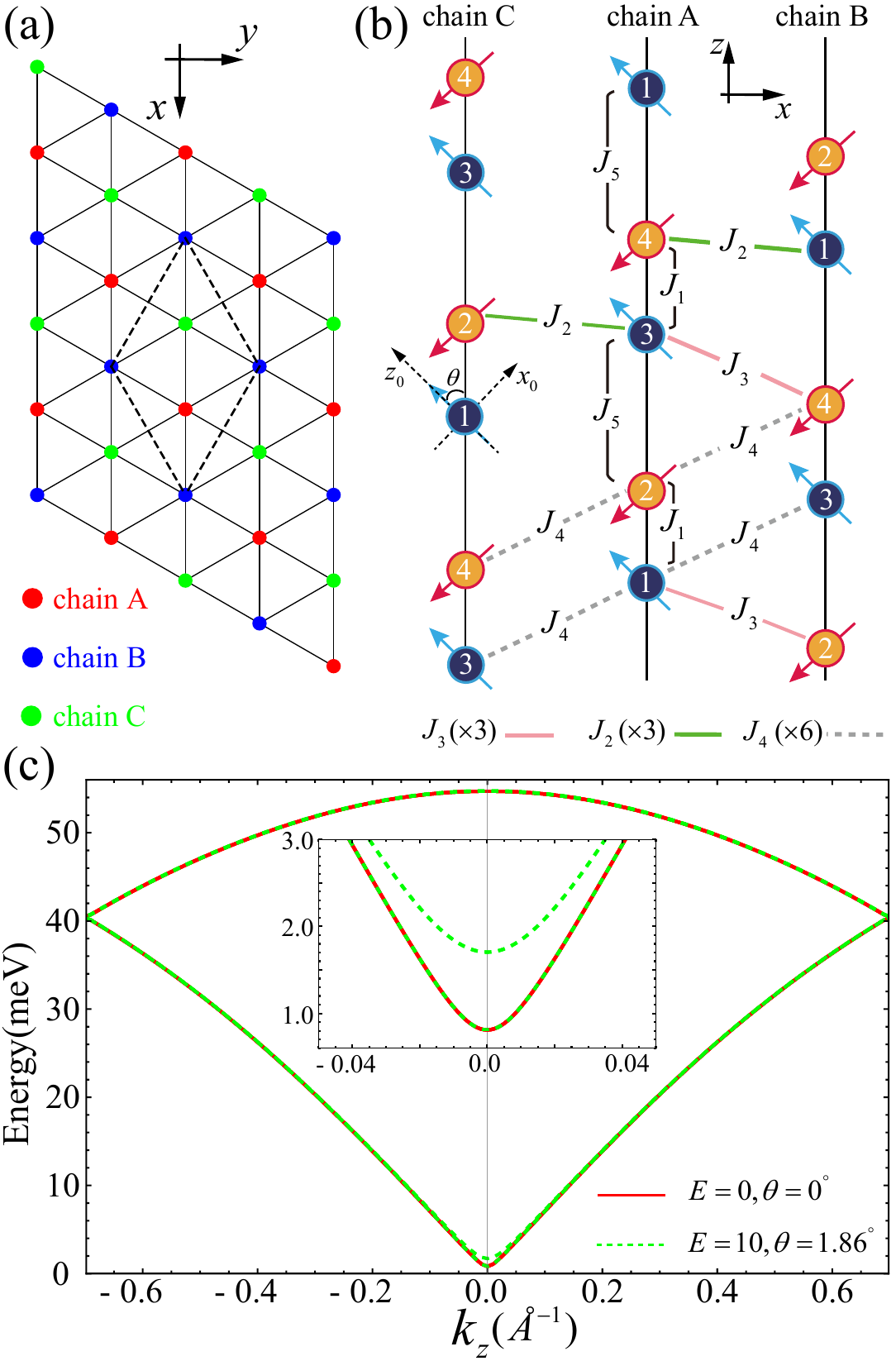}
	\caption{
		The crystal structure and magnon dispersion of Cr$_2$O$_3$.
		(a) The lattice structure viewed from the $z$ axis. The $A$, $B$, and $C$ chains correspond to the respective chains in (a). The dashed rhombus shows the unit cell of Cr$_2$O$_3$.
		(b) A schematic of the crystal structure and exchange interactions $J_1$-$J_5$ viewed from the $y$ axis. Numbers in the parentheses indicate the number of bonds.
		(c) Magnon dispersion for the model in Eq.~\eqref{eq:3dmodel} with $E=0$ (red) and $E=10$ (dashed green), where the unit of the product $\lambda E$ is meV and $\lambda = 0.05$. The results are for $J_1 = -7.53$ meV, $J_2 = -3.41$ meV, $J_3 =-0.08$ meV, $J_4 = 0.02$ meV, $J_5 = -0.19$ meV~\cite{Samuelsen1970a}, and $D_z = 0.0015$ meV~\cite{Foner1963a, Mu2019a}. Enlarged views of the lower magnon band near the $\Gamma$ point are shown in the inset.
	}\label{fig:3D_portrait}
\end{figure}

To assess the experimental relevance, we next consider a three-dimensional spin model relevant to chromium sesquioxide (Cr$_2$O$_3$)~\cite{Samuelsen1970a}.
The crystal and magnetic structure of Cr$_2$O$_3$ are shown in Fig.~\ref{fig:3D_portrait}. 
Cr$_2$O$_3$ is an antiferromagnetic insulator with a trigonal corundum structure (space group $R\bar{3}c$)~\cite{Schober1995a,Hill2010a}. The primitive cell contains four Cr ions aligned along the $z$ axis, which aligns in $\uparrow \downarrow \uparrow \downarrow$ antiferromagnetic configuration at the low temperatures~\cite{Corliss1965a,Catti1996a}; the magnetic order breaks inversion symmetry and gives rise to the magnetoelectric effects~\cite{Dyaloshinskii1960a,Astrov1961a,Rado1961a,Iyama2013a}.

The exchange interactions in Cr$_2$O$_3$ have been studied by neutron scattering experiments and first-principles calculations~\cite{Samuelsen1970a,Shi2009a}, which are shown in Fig.~\ref{fig:3D_portrait}.
Building on these studies, we consider a model with five exchange interactions (Fig.~\ref{fig:3D_portrait}(b)); $J_1$ and $J_5$ are intrachain couplings, and $J_2$, $J_3$, and $J_4$ account for interchain interactions with different coordination numbers.
The Hamiltonian reads,
\begin{align}
	\mathcal{H}^{3D} = \mathcal{H}_{1}^{3D} + \mathcal{H}_{2}^{3D} +\mathcal{H}_{\text{DMI}}^{3D} + \mathcal{H}_D,\label{eq:3dmodel}
\end{align}
where $\mathcal{H}_D$ is the same as that in Eq.~\eqref{eq:1D_Hamiltonian_initial}, and the other three terms are,
\begin{align}
	\mathcal{H}_{1}^{3D} =   & - \sum\limits_{n} \sum\limits_j J_1{{\bm S}_{n,j,1}} \cdot {{\bm S}_{n,j,2}} + J_1{{\bm S}_{n,j,3}} \cdot {{\bm S}_{n,j,4}}\nonumber \\
	                         & \qquad\qquad+ J_5{{\bm S}_{n,j,2}} \cdot {{\bm S}_{n,j,3}} + J_5{{\bm S}_{n,j,4}} \cdot {{\bm S}_{n,j+1,1}}, \nonumber               \\
	\mathcal{H}_{2}^{3D} =   & - \sum\limits_{\langle n,m \rangle
	} \sum\limits_j J_3 ({{\bm S}_{n,j,1}} \cdot {{\bm S}_{m,j,2}} + {{\bm S}_{n,j,3}} \cdot {{\bm S}_{m,j,4}})\nonumber                                            \\
	                         & \qquad\qquad+ J_2({{\bm S}_{n,j,2}} \cdot {{\bm S}_{m,j,3}} + {{\bm S}_{n,j,4}} \cdot {{\bm S}_{m,j+1,1}}) \nonumber                 \\
	                         & \qquad\qquad+ J_4 ({{\bm S}_{n,j,1}} \cdot {{\bm S}_{m,j,3}} + {{\bm S}_{n,j,2}} \cdot {{\bm S}_{m,j,4}}), \nonumber                 \\
	\mathcal{H}_{DMI}^{3D} = & \bm{\mathcal{D}} \cdot \sum_n \sum_{j} \left(\bm{S}_{n,j,1} \times \bm{S}_{n,j,2} + \bm{S}_{n,j,3} \times \bm{S}_{n,j,4}\right).
\end{align}

The magnon bands of Eq.~\eqref{eq:3dmodel} are plotted in Fig.~\ref{fig:3D_portrait}(c). Similar to the one-dimensional model, the magnon bands are doubly degenerate at $E = 0$, whereas a finite electric field $E$ lifts the degeneracy, as shown in the inset of Fig.~\ref{fig:3D_portrait}(c).

\begin{figure}
	\includegraphics[width=\linewidth]{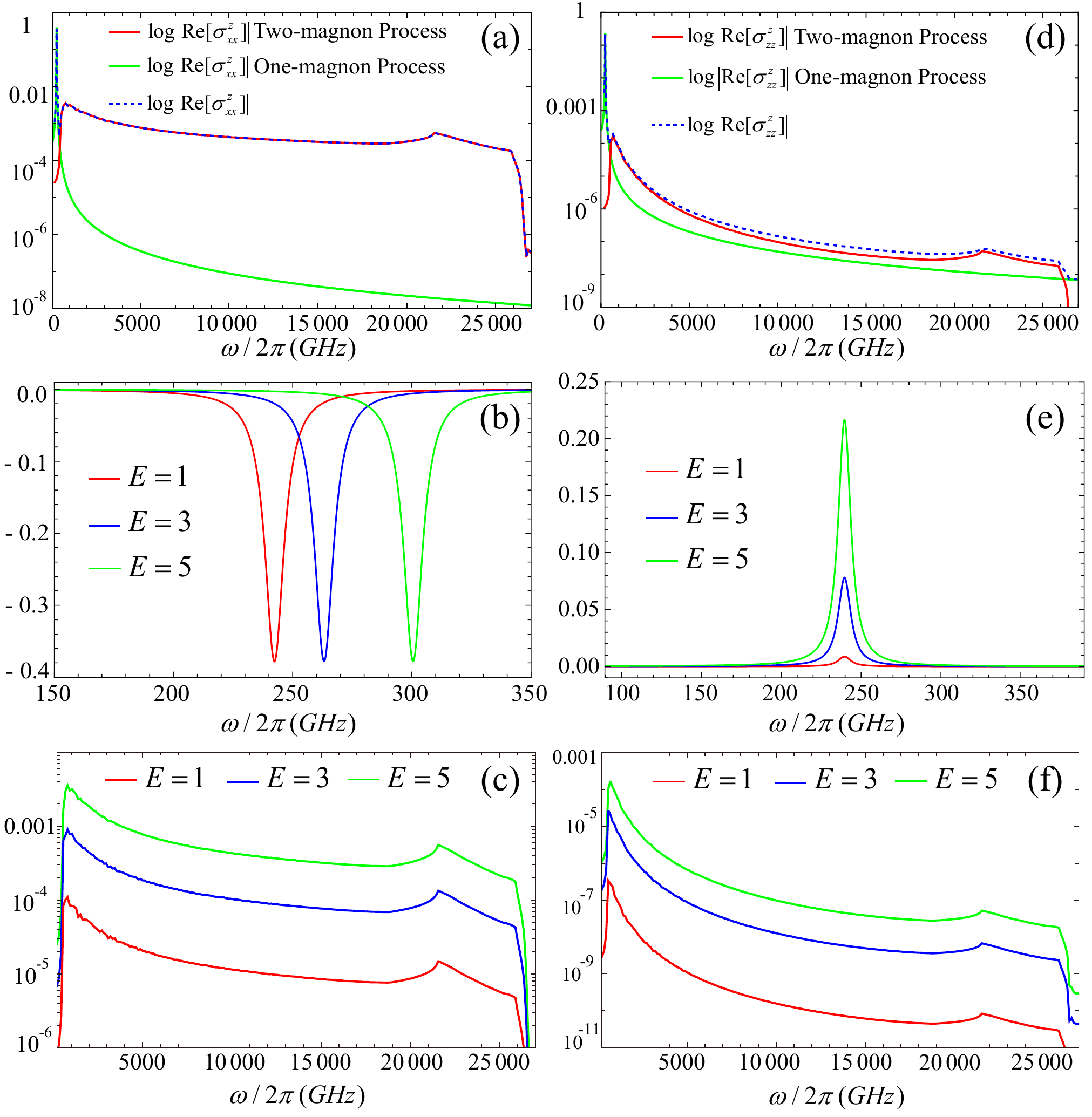}
	\caption{
		Nonlinear spin current conductivities $\sigma_{xx}^z(0;\omega,-\omega)$ and $\sigma_{zz}^z(0;\omega,-\omega)$. (a) Frequency dependence of $[\sigma^{(2)}]_{xx}^z$and its (b) one-magnon and (c) two-magnon contributions. (d) Frequency dependence of $[\sigma^{(2)}]_{zz}^z$ and (e) the one-magnon and (f) the two-magnon processes. The canting angle $\theta$ for $ E = 1$, 3, and 5 are $\theta = 0.09^\circ$, $0.24^\circ$, and $0.39^\circ$, respectively. The results are for  Gilbert damping $\alpha = 0.02$, $J_1=-7.53$ meV, $J_2=-3.41$ meV, $J_3=-0.08$ meV, $J_4=0.02$ meV, $J_5=-0.19$ meV, $D_z=0.0015$ meV, $\eta=0.02E_1(\mathbf{k}=0)$ meV, and $\lambda = 0.05$. The unit of product $\lambda E$ is energy scale (meV).
	}\label{fig:3D_conductivity}
\end{figure}

\begin{figure}
	\includegraphics[width=\linewidth]{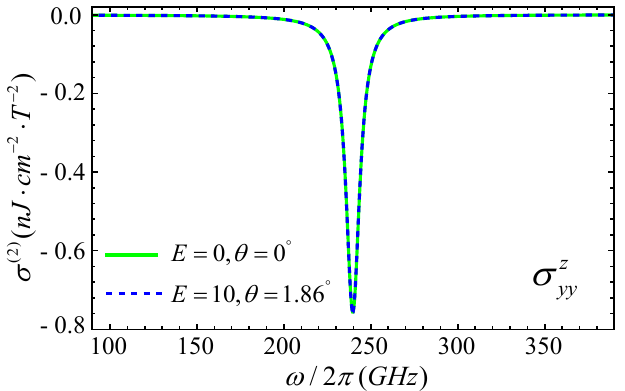}
	\caption{
		Nonlinear spin current conductivity $\sigma_{yy}^z(0;\omega,-\omega)$ with Gilbert damping $\alpha = 0.02$, $J_1 = - 7.53$ meV, $J_2 = -3.41$ meV, $J_3 = -0.08$ meV, $J_4 = 0.02$ meV, $J_5 = -0.19$ meV, $D_z = 0.0015$ meV, and $\eta = 0.02\omega_1(\bm k=0)$ meV. The unit of $\lambda E$ is meV.
	}\label{fig:3D_conductivity_y}
\end{figure}

The frequency dependence of $[\sigma^{(2)}]_{xx}^z(0;\omega,-\omega)$ and $[\sigma^{(2)}]_{zz}^z(0;\omega,-\omega)$ are shown in Figs.~\ref{fig:3D_conductivity}(a)--(c) and (d)--(f), respectively. Here, the magnon relaxation rate is taken as $\eta=\alpha E_1(k=0)\approx 4.79$ meV~\cite{Abrikosov1963a}, where $\alpha=0.02$ is the damping constant reported experimentally for Cr$_2$O$_3$~\cite{Nguyen2020a}. Moreover, we adopted $\lambda = 0.05$ and $E = 1$--$5$ as dimensionless fitting parameters, such that the product $\lambda E = 0.05 \sim 0.25$ meV sets the relevant energy scale used in Figs.~\ref{fig:3D_conductivity} and \ref{fig:3D_conductivity_y}.

In Figs.~\ref{fig:3D_conductivity}(a) and (d), we present the conductivity in the range from 200 GHz to 27 THz, corresponding to twice the Brillouin-zone bandwidth. This frequency range is consistent with the width of the continuum spectrum of the two-magnon process discussed previously for the 1D case.
The dashed blue curves denote the total conductivity, while the green and red curves represent the one-magnon and two-magnon contributions, respectively. In Fig.~\ref{fig:3D_conductivity}(a), the total conductivity exhibits a clear secondary peak, originating from the superposition of the broad two-magnon continuum with the sharp one-magnon resonance. This feature is more pronounced in $[\sigma^{(2)}]_{xx}^z$, where the two-magnon contribution is larger and lies closer to the one-magnon resonance peak.

Furthermore, the electric-field dependence of the one-magnon process in Figs.~\ref{fig:3D_conductivity}(b) and \ref{fig:3D_conductivity}(e), as well as that of the two-magnon process in Figs.~\ref{fig:3D_conductivity}(c) and \ref{fig:3D_conductivity}(f), follows the same trend as in the 1D model discussed in Sec.~\ref{sec:1D}. A notable difference, however, is that no clear multi-peak structure appears in the two-magnon contribution of $[\sigma^{(2)}]_{xx}^z$ in Fig.~\ref{fig:3D_conductivity}(c).
Specifically, the two-magnon peaks remain broad and unsplit within each frequency range, rather than developing the multi-peak continuum structure observed in Fig.~\ref{fig:1D_conductivity_twoprocess}(a).

The absence of multiple peaks is a consequence of two factors: the canting angle $\theta$ adopted in the 3D model is smaller than that in the 1D model, and the absence of a singularity at the edge of the magnon DOS.
Firstly, the canting angle $\theta$ for the same $E$ is smaller in the 3D case. The small $\theta$ results from the different values of the exchange parameters $J_i$ and from the fact that more exchange interactions are included in the 3D model, both of which reduce $\theta$ through Eq.~\eqref{eq:models_theta}. Secondly, the two-magnon contribution to the conductivity in the 3D model is given by
\begin{align}
	\mathrm{Re}\left[\sigma_{\mu\nu\lambda}^{(2)}\right]
	\propto
	\sum_{nml}
	\int_{\Delta_{nm}(\bm k)=\hbar\omega}
	\frac{
		dS_{\bm k}\mathcal W_{nml}(\bm k)
	}
	{
		|\nabla_{\bm k}\Delta_{mn}(\bm k)|
	},
\end{align}
which reflects the JDOS.
Near $\bm k = 0$, the JDOS is proportional to $1/\sqrt{\hbar \omega - \Delta_{mn}(0)}$ in the 1D case, whereas it is proportional to $\sqrt{\hbar \omega - \Delta_{mn}(\bm 0)}$ in the 3D case, where $\Delta_{mn}(\bm 0) = \gamma_m E_{m\bm 0} - \gamma_n E_{n\bm0}$.
The former gives rise to a van Hove singularity for JDOS in the low-frequency region, whereas the latter does not.
Consequently, a sharp peak appears in the 1D case but not in the 3D counterpart.
The combined effect of the reduced canting angle and the absence of the JDOS singularity gives rise to a different spectrum.

Figure~\ref{fig:3D_conductivity_y} shows the frequency dependence of $[\sigma^{(2)}]_{yy}^z(0;\omega,-\omega)$. For the light polarized along the $y$ direction, the two-magnon process is absent, as in Eq.~\eqref{eq:model_perturbation_y}.
This absence is directly related to the spin rotation in Eq.~\eqref{eq:method_rotation_transformation}.
\begin{table*}
	\centering
	\setlength{\tabcolsep}{6pt}
	\renewcommand{\arraystretch}{1.15}

	\begin{tabular}{llcccc}
		\hline\hline
		Type of mechanism & Component               & $\sigma$ (J$\cdot$cm$^{-2}$$\cdot$T$^{-2}$) & $B$ (mT) & $\omega$ (GHz) \\
			\hline

			Collinear (One-Magnon process)
		                  & $\sigma_{xx}^z$
		                  & $10^{-10}\!-\!10^{-9}$
		                  & $0.1\!-\!1$
		                  & $240$                                                                                             \\

			\hline

			\multirow{3}{*}{Noncollinear (One-Magnon process)}
		                  & $\sigma_{xx}^{z}$
		                  & $10^{-10}\!-\!10^{-9}$
		                  & $0.1\!-\!1$
		                  & $240\!-\!270$                                                                                     \\
		                  & $\sigma_{yy}^{z}$
		                  & $10^{-10}\!-\!10^{-9}$
		                  & $0.1\!-\!1$
		                  & $240$                                                                                             \\
		                  & $\sigma_{zz}^{z}$
		                  & $10^{-11}\!-\!10^{-10}$
		                  & $1\!-\!10$
		                  & $240$                                                                                             \\

			\hline

			\multirow{2}{*}{Noncollinear (Two-Magnon process)}
		                  & $\sigma_{zz}^{z}$
		                  & $10^{-15}\!-\!10^{-13}$
		                  & $>100$
		                  & $500$                                                                                             \\
		                  & $\sigma_{xx}^{z}$
		                  & $10^{-13}\!-\!10^{-12}$
		                  & $10 - 100$
		                  & $500\!-\!600$                                                                                     \\

		\hline\hline
	\end{tabular}
	\caption{Estimated experimental requirements for detecting the predicted conductivity components.Here $B$ denotes the amplitude of the ac magnetic field of the incident light.}
	\label{tab:tab_3D_estimate_results_conductivity}
\end{table*}

Since the ordered moments are rotated in the $x-z$ plane, the $y$ axis is the rotation axis and therefore remains unchanged under the transformation, i.e., $S_j^y=S_j^{y_0}$.
Consequently, the $y$-polarized perturbation contains only the transverse spin component in the local frame, which gives rise to the one-magnon process.
In contrast, the $x$ and $z$ components are mixtures of $S_j^{x_0}$ and $S_j^{z_0}$ after the rotation.
The longitudinal component $S_j^{z_0}$ contains quadratic bosonic terms and thus provides the origin of the two-magnon contribution.
Consequently, $[\sigma^{(2)}]_{yy}^z(0;\omega,-\omega)$ exhibits only a resonance peak related to the single one-magnon process. The resonance frequency depends only weakly on the electric field $E$, reflecting the limited modification of the magnon spectrum.

Having established the nonlinear response of the realistic three-dimensional model, we now discuss the experimental feasibility of observing this effect. Based on previous experimental data~\cite{Astrov1961a,Folen1961a,Royce1963a,Artman1962a} and the analysis by Hornreich~\cite{Hornreich1967a}, the magnetoelectric coupling constant is estimated to be $\lambda = (2.61$--$6.64)\times10^{-9}\,\mathrm{meV}\,\cdot\,\mathrm{m}\,\cdot\,\mathrm{V}^{-1}$.
Hence, the required electric field for $\lambda E =0.05$--$0.15$ meV used in Fig.~\ref{fig:3D_conductivity} is approximately $10^7$--$10^8$ V/m.
It is somewhat above the typical voltage used in the experiments~\cite{Yokota2006a,Ashida2014a,Toyoki2015a}, but still below the electric breakdown of Cr$_2$O$_3$~\cite{Sun2017a}.
Hence, a noticeable shift of the resonance peaks in the spin current should be observable.

We next estimate the experimentally detectable magnitude of the nonlinear spin photocurrent. Previous studies have suggested that spin current larger than $J>10^{-16} \, \mathrm{nJ} \cdot \mathrm{cm}^{-2}$ can be detected experimentally~\cite{Ishizuka2019a,Ishizuka2022a,Gu2025a}.
Using this criterion, the required magnetic-field amplitude of the incident laser is estimated and shown in Table~\ref{tab:tab_3D_estimate_results_conductivity}.
These results indicate that the predicted nonlinear spin-current response in Cr$_2$O$_3$ can be observable in experiment, including the two-magnon process $\sigma_{xx}^z$.
In addition, the anisotropic response suggests potential applications in polarization-selective spin photodetectors and electrically tunable optospintronic devices.

\section{Summary and Discussions}\label{sec:discussion}

In this work, we investigated the nonlinear magnon spin photocurrent in the antiferromagnet Cr$_2$O$_3$ using nonlinear response theory, based on a four-sublattice spin model. We first studied a one-dimensional model
and found that the spin-current conductivity exhibits pronounced anisotropy with respect to the polarization direction of the incident light; both the resonance frequency and the peak intensity depend sensitively on the external electric field $E$, demonstrating electrical
control of the photoinduced spin current. Extending the analysis to a realistic three-dimensional model of Cr$_2$O$_3$, we found that this anisotropic nonlinear response remains robust for light polarized along
each direction $\alpha = x, y, z$. Such directional dependence provides a useful signature for distinguishing the magnon spin photocurrent from competing mechanisms, such as inhomogeneous thermal effects. Drawing on
available experimental data, we further estimated realistic values of the magnetoelectric coupling $\lambda$, the electric field $E$, and the optical intensity $B$ required for detection. These results establish Cr$_2$O$_3$ as a promising platform for electrically tunable nonlinear spin photocurrents and polarization-selective antiferromagnetic optospintronic applications.

Finally, we comment on the role of magnon--phonon coupling, which has been experimentally identified in the optical phonon modes of Cr$_2$O$_3$~\cite{Sun2025a,Li2020a}. In Cr$_2$O$_3$, the phonon and magnon dispersions partially overlap in energy, especially for the high-energy branch near the $\Gamma$ point~\cite{Schober1995a,Sun2025a}. For the one-magnon contribution, the spin photocurrent originates exclusively from the acoustic magnon mode at $\bm{k}=\bm{0}$, as given in Eq.~\eqref{eq:model_definition_spin_conductivity_one}. Hence, the interaction between this low-energy magnon excitation and a high-energy polar optical phonon is expected to be negligible due to the large energy mismatch, in agreement with previous work~\cite{Gu2025a}. For the two-magnon contribution, on the other hand, magnon--phonon coupling may modify the continuum spectrum in the high-frequency regime.
Studies on such effects are left for future research.

\acknowledgements
We are grateful to Y. Liu, H. Murata, and H. Yoshida for fruitful discussions.
This work is supported by JSPS KAKENHI (Grant No. JP19K14649, No. JP23K03275, and No. JP25H00841), JST PRESTO (Grant No. JPMJPR2452), and JST SPRING (Grant No. JPMJSP2180).

\appendix

\section{Derivation of conductivity for the two-magnon process}
In this appendix, we derive the second-order response formula, Eq.~\eqref{eq:model_definition_spin_conductivity_two}, used in the main text for the two-magnon process in a general quadratic bosonic system. The final expression is written in a form suitable for an $N$-mode bosonic Bogoliubov--de Gennes Hamiltonian, which is a generalization of the two-band model investigated in a previous work~\cite{Ishizuka2019a}.

We consider a time-dependent Hamiltonian,
\begin{align}
	H(t) = H_0 + H'(t),
\end{align}
where the perturbation operator is written as
\begin{align}
	H'(t) = -\sum_{\mu} F_{\mu}(t)\hat{B}_{\mu}.
\end{align}
Here, $\hat{B}_{\mu}$ is the corresponding coupling operator coupled to an external field $F_{\mu}(t)$. The expectation value of an observable $\hat A$ at time $t$ is given by
\begin{align}
	\langle \hat{A} \rangle (t) = \frac{\mathrm{Tr}\left[\rho(t)\hat{A}\right]}{\mathrm{Tr}\left[\rho(t)\right]},
\end{align}
with $\rho(t)$ being the density matrix.

To obtain the nonlinear response, we expand $\rho(t)$ to second order in the external fields. After the Fourier transformation, the response takes the form
\begin{align}
	\langle A \rangle (\Omega)
	= \sum_{\mu,\nu} \int d\omega \;
	\sigma_{\mu\nu}(\Omega;\omega,\Omega-\omega)\,
	F_\mu(\omega)\,F_\nu(\Omega-\omega).
\end{align}
In the eigenbasis of $H_0$, the second-order conductivity is expressed as
\label{sec:appendix_derive_conductivity_TMP}
\begin{widetext}
	\begin{align}\label{eq:app_general_kubo}
		\sigma_{\mu \nu}^{(2)}(\Omega;\omega,\Omega - \omega) = \frac{1}{2\pi}
		\sum_{n,m,l}
		\frac{\rho_n - \rho_m}{\hbar \omega + E_n - E_m - i\hbar \eta}
		\left(
		\frac{B_{nm}^\mu B_{ml}^\nu A_{ln}}{\hbar \Omega + E_n - E_l - i\hbar \eta}
		-
		\frac{B_{nm}^\mu A_{ml} B_{ln}^\nu}{\hbar \Omega + E_l - E_m - i\hbar \eta}
		\right),
	\end{align}
	$E_n$ and $\rho_n$ are the energy and occupation probability of the many-body eigenstate $|n\rangle$, respectively. The parameter $\eta = 1/2\tau$ phenomenologically describes relaxation, where $\tau$ is the relaxation time.

	We next specialize Eq.~\eqref{eq:app_general_kubo} to a quadratic bosonic Hamiltonian. After the Bogoliubov transformation, the unperturbed Hamiltonian is diagonalized as
	\begin{align}
		H_0 = \sum_{n=1}^{N} \hbar \omega_n b_n^\dagger b_n.
	\end{align}
	For later convenience, we introduce a $2N$-component Nambu operator
	\begin{align}
		\psi_n =
		\begin{cases}
			b_n,             & 1\le n\le N,    \\[4pt]
			b_{n-N}^\dagger, & N+1\le n\le 2N,
		\end{cases}
	\end{align}
	and define
	\begin{align}
		\gamma_n =
		\begin{cases}
			+1, & 1\le n\le N,     \\[4pt]
			-1, & N+1\le n\le 2N .
		\end{cases}
	\end{align}
	With this convention, $\gamma_n\omega_n$ denotes the signed quasiparticle energy, where $\omega_{n+N}\equiv \omega_n$.

	The observable and perturbation operators are written as quadratic forms in the Nambu basis,
	\begin{align}\label{eq:app_quadratic_operator}
		\hat A
		=
		\frac{1}{2}
		\sum_{n,m=1}^{2N}
		a_{nm}\,
		\psi_n^\dagger\psi_m ,
		\qquad
		\hat B^\mu
		=
		\frac{1}{2}
		\sum_{n,m=1}^{2N}
		\beta^\mu_{nm}\,
		\psi_n^\dagger\psi_m .
	\end{align}
	Equivalently, Eq.~\eqref{eq:app_quadratic_operator} contains all number-conserving and anomalous terms,
	\begin{align}\label{eq:app_A_operator}
		\hat A
		=
		\frac{1}{2}\sum_{n,m=1}^{N}
		\left(
		a_{nm} b_n^\dagger b_m
		+a_{n,m+N} b_n^\dagger b_m^\dagger
		+a_{n+N,m} b_n b_m
		+a_{n+N,m+N} b_n b_m^\dagger
		\right),
	\end{align}
	and similarly
	\begin{align}\label{eq:app_B_operator}
		\hat B^\mu
		=
		\frac{1}{2}\sum_{n,m=1}^{N}
		\left(
		\beta^\mu_{nm} b_n^\dagger b_m
		+\beta^\mu_{n,m+N} b_n^\dagger b_m^\dagger
		+\beta^\mu_{n+N,m} b_n b_m
		+\beta^\mu_{n+N,m+N} b_n b_m^\dagger
		\right).
	\end{align}
	This form is the natural generalization of the two-band expression~\cite{Ishizuka2019a} to an $N$-mode bosonic system.

	Substituting the quadratic operators into Eq.~\eqref{eq:app_general_kubo}, the conductivity can be written as
	\begin{align}\label{eq:app_six_operator}
		\sigma_{\mu \nu}^{(2)}(\Omega;\omega,\Omega - \omega) & = \frac{1}{16\pi \hbar^2} \sum_{k_i=1}^{2N} \frac{1 - e^{-\beta_T (\gamma_{k_2} \omega_{k_2} - \gamma_{k_1} \omega_{k_1})}}
		{\omega + \gamma_{k_1} \omega_{k_1} - \gamma_{k_2} \omega_{k_2} - i\eta}
		\left(
		\frac{\beta_{k_1 k_2}^\mu \beta_{k_3 k_4}^\nu a_{k_5 k_6}}
		{\Omega + \gamma_{k_6} \omega_{k_6} - \gamma_{k_5} \omega_{k_5} - i\eta}
		-
		\frac{\beta_{k_1 k_2}^\mu a_{k_3 k_4} \beta_{k_5 k_6}^\nu}
		{\Omega + \gamma_{k_4} \omega_{k_4} - \gamma_{k_3} \omega_{k_3} - i\eta}
		\right) \nonumber                                                                                                                                                                   \\
		                                                      & \quad \times \langle \psi_{k_1}^\dagger \psi_{k_2} \psi_{k_3}^\dagger \psi_{k_4} \psi_{k_5}^\dagger \psi_{k_6} \rangle_0,
	\end{align}
	where $\beta_T = 1/(k_B T)$ is the thermal temperature and $k_B$ is the Boltzmann constant. Here, $\langle \cdots \rangle_0$ denotes the thermal average with respect to $H_0$. While the operators are quadratic in bosonic quasiparticles, the six-point function in Eq.~\eqref{eq:app_six_operator} can be reduced by Wick's theorem. Only contractions compatible with the Nambu structure survive, including $\psi_n^\dagger \psi_n$, $\psi_{n+N}\psi_{n+N}^\dagger$, $\psi_n^\dagger \psi_{n+N}^\dagger$, $\psi_n \psi_{n+N}$. Using
	\begin{align}
		n_n=\frac{1}{e^{\beta_T\hbar\omega_n}-1},
	\end{align}
	we find
	\begin{align}\label{eq:app_wick_result}
		 & \left\langle
		\psi_{k_1}^\dagger\psi_{k_2}
		\psi_{k_3}^\dagger\psi_{k_4}
		\psi_{k_5}^\dagger\psi_{k_6}
		\right\rangle_0
		=
		\left(
		n_{k_1}+\frac{1-\gamma_{k_1}}{2}
		\right)
		\left(
		n_{k_2}+\frac{1+\gamma_{k_2}}{2}
		\right) \Big[
		\gamma_{k_4}\delta_{k_1+\gamma_{k_1}N,k_3}\delta_{k_2k_5}\delta_{k_4+\gamma_{k_4}N,k_6} \nonumber \\
		 & \hspace{3cm}
			+ \gamma_{k_4}\delta_{k_1+\gamma_{k_1}N,k_3}\delta_{k_2+\gamma_{k_2}N,k_6}\delta_{k_4k_5}
			-\gamma_{k_3}\delta_{k_1k_4}\delta_{k_2k_5}\delta_{k_3k_6}
			-\gamma_{k_3}\delta_{k_1k_4}\delta_{k_2+\gamma_{k_2}N,k_6}\delta_{k_3+\gamma_{k_3}N,k_5}
			\Big].
	\end{align}
	The thermal prefactor in Eqs.~\eqref{eq:app_six_operator} and~\eqref{eq:app_wick_result} can be simplified as
	\begin{align}\label{eq:app_thermal_identity}
		\left[
		1-e^{-\beta_T\hbar(\gamma_m\omega_m-\gamma_n\omega_n)}
		\right]
		\left(
		n_n+\frac{1-\gamma_n}{2}
		\right)
		\left(
		n_m+\frac{1+\gamma_m}{2}
		\right) =
		\gamma_m
		\left(
		n_n+\frac{1-\gamma_n}{2}
		\right)
		-
		\gamma_n
		\left(
		n_m+\frac{1-\gamma_m}{2}
		\right).
	\end{align}
	After carrying out the Kronecker-delta sums, the conductivity becomes
	\begin{align}\label{eq:app_finite_T_conductivity}
		\sigma_{\mu\nu\lambda}^{(2)}(\Omega; \omega, \Omega - \omega) =
		\frac{1}{16\pi \hbar^2} \sum_{n,m,l=1}^{2N}
		\gamma_l \frac{\gamma_m \left( n_n + \frac{1 - \gamma_n}{2} \right) - \gamma_n \left( n_m + \frac{1 - \gamma_m}{2} \right)}
		{\omega + \gamma_n \omega_n - \gamma_m \omega_m - i\eta}
		\left[
			\frac{\beta_{nm}^\mu \beta_{n+\gamma_n N, l}^\nu \left( a^\lambda_{m,l+\gamma_l N} + a^\lambda_{l,m+\gamma_m N} \right)}
			{\Omega - \gamma_l \omega_l - \gamma_m \omega_m - i\eta} \right.
		         & \nonumber \\
			\left.
			- \frac{\beta_{nm}^\mu a^\lambda_{n+\gamma_n N, l} \left( \beta_{m,l+\gamma_l N}^\nu + \beta_{l,m+\gamma_m N}^\nu \right)}
			{\Omega + \gamma_l \omega_l + \gamma_n \omega_n - i\eta}
			- \frac{\beta_{nm}^\mu \beta_{l n}^\nu \left( a^\lambda_{m l} + a^\lambda_{l+\gamma_l N, m+\gamma_m N} \right)}
			{\Omega + \gamma_l \omega_l - \gamma_m \omega_m - i\eta}
			+ \frac{\beta_{nm}^\mu a^\lambda_{l n} \left( \beta_{m l}^\nu + \beta_{l+\gamma_l N, m+\gamma_m N}^\nu \right)}
			{\Omega + \gamma_n \omega_n - \gamma_l \omega_l - i\eta}
		\right]. &
	\end{align}

	We finally take the zero-temperature limit, which is the regime mainly considered in this work. Since $n_n = n_m = 0$ for all positive-energy bosonic modes, Eq.~\eqref{eq:app_finite_T_conductivity} reduces to
	\begin{align}
		 & \sigma_{\mu\nu\lambda}^{(2)}(\Omega; \omega, \Omega - \omega) =
		\frac{1}{32\pi \hbar^2} \sum_{n,m,l=1}^{2N}
		\frac{\gamma_l(\gamma_m  - \gamma_n)}
		{\omega + \gamma_n \omega_n - \gamma_m \omega_m - i\eta}
		\left[
			\frac{\beta_{nm}^\mu \beta_{n+\gamma_n N, l}^\nu \left( a^\lambda_{m,l+\gamma_l N} + a^\lambda_{l,m+\gamma_m N} \right)}
			{\Omega - \gamma_l \omega_l - \gamma_m \omega_m - i\eta} \right.
		\nonumber                                                          \\
		 & \left.
			- \frac{\beta_{nm}^\mu a^\lambda_{n+\gamma_n N, l} \left( \beta_{m,l+\gamma_l N}^\nu + \beta_{l,m+\gamma_m N}^\nu \right)}
			{\Omega + \gamma_l \omega_l + \gamma_n \omega_n - i\eta}
			- \frac{\beta_{nm}^\mu \beta_{l n}^\nu \left( a^\lambda_{m l} + a^\lambda_{l+\gamma_l N, m+\gamma_m N} \right)}
			{\Omega + \gamma_l \omega_l - \gamma_m \omega_m - i\eta}
			+ \frac{\beta_{nm}^\mu a^\lambda_{l n} \left( \beta_{m l}^\nu + \beta_{l+\gamma_l N, m+\gamma_m N}^\nu \right)}
			{\Omega + \gamma_n \omega_n - \gamma_l \omega_l - i\eta}
			\right].
	\end{align}
	This is the formula used in the main text. Although no thermally excited magnons are present at $T=0$, the response does not necessarily vanish. The nonzero contribution originates from the anomalous matrix elements in Eqs.~\eqref{eq:app_A_operator} and \eqref{eq:app_B_operator}, which describe the creation and annihilation of magnon pairs. Thus, the two-magnon process plays the same technical role as the interband transition in ordinary shift-current formulas, while the present expression is written for a general $N$-mode bosonic system.
\end{widetext}

\section{Relation between Conductivity and Density of States}
\label{sec:appendix_conductivity_DOS}

In this appendix, we discuss the relation between the spin conductivity and the density of states for the two-magnon process. By using the identity
\begin{align}
	\frac{1}{x-i\eta}
	=
	\mathcal P\frac{1}{x}
	+ i\pi\delta(x),
\end{align}
where $\mathcal P$ denotes the Cauchy principal value, the resonant contribution to Eq.~\eqref{eq:model_definition_spin_conductivity_two} can be expressed as
\begin{align}
	 & \mathrm{Re}\left[\sigma_{\mu\nu\lambda}^{(2)}\right]
	\propto \nonumber                                       \\
	 & - \sum_{nml} \int dk\,
	\mathcal \gamma_l F_{nm}\,
	\mathrm{Im}\!\left[
		\mathcal M_{\mu\nu\lambda}^{nml}(\bm k)
		\right]
	\delta\!\left[
		\hbar\omega-\Delta_{mn}(\bm k)
		\right],
\end{align}
where the principal-value contribution has been omitted, and
\begin{align}
	\mathcal F_{nm}
	 & =
	\gamma_m - \gamma_n, \\
	\Delta_{mn}(\bm k)
	 & =
	\gamma_m E_{m \bm k}-\gamma_n E_{n \bm k}.
\end{align}
Here, $\mathcal M_{\mu\nu\lambda}^{nml}(k)$ is defined as
\begin{widetext}
	\begin{align}
		\mathcal{M}_{\mu\nu\lambda}^{nml}(\bm k) & =
		\frac{[\tilde{\beta}_{\bm k}^\nu]_{nm} [\tilde{\beta}_{\bm k}^\lambda]_{n+\gamma_n N, l} \left([{\tilde{\mathcal{J}}}_{\bm k}^\mu]_{m,l+\gamma_l N} + [{\tilde{\mathcal{J}}}_{\bm k}^\mu]_{l,m+\gamma_m N} \right)}
		{ - \gamma_l E_{l\bm k} - \gamma_m E_{m\bm k} - i\eta}
		-\frac{[\tilde{\beta}_{\bm k}^\nu]_{nm} [{\tilde{\beta}}_{\bm k}^\lambda]_{l n} \left( [\tilde{\mathcal{J}}_{\bm k}^\mu]_{m l} + [\tilde{\mathcal{J}}_{\bm k}^\mu]_{l+\gamma_l N, m+\gamma_m N} \right)}
		{\gamma_l E_{l\bm k} - \gamma_m E_{m\bm k} - i\eta}
		\nonumber                                                                                                                                                                                                                                                      \\
		                                         & - \frac{[\tilde{\beta}_{\bm k}^\nu]_{nm} {[\tilde{\mathcal{J}}_{\bm k}^{\mu}]}_{n+\gamma_n N, l} \left([\tilde{\beta}_{\bm k}^\lambda]_{m,l+\gamma_l N} + [\tilde{\beta}_{\bm k}^\lambda]_{l,m+\gamma_m N} \right)}
		{\gamma_l E_{l\bm k} + \gamma_n E_{n\bm k} - i\eta}
		+ \frac{[\tilde{\beta}_{\bm k}^\nu]_{nm} [\tilde{\mathcal{J}}_{\bm k}^\mu]_{l n} \left([\tilde{\beta}_{\bm k}^{\lambda}]_{m l} +[\tilde{\beta}_{\bm k}^\lambda]_{l+\gamma_l N, m+\gamma_m N} \right)}
		{\gamma_n E_{n\bm k} - \gamma_l E_{l\bm k} - i\eta}.
	\end{align}
\end{widetext}

Using the property of the delta function in 1D,
\begin{align}
	\delta[f(k)]
	=
	\sum_{k_i}
	\frac{\delta(k-k_i)}
	{|f'(k_i)|},
\end{align}
where $k_i$ satisfies
$\hbar\omega-\Delta_{mn}(k_i)=0$, we obtain
\begin{align}\label{eq:appendix_conductivity}
	\mathrm{Re}\left[\sigma_{\mu\nu\lambda}^{(2)}\right]
	\propto
	\sum_{nml}
	\sum_{k_i}
	\frac{
	\mathcal W_{nml}(k_i)
	}
	{
	\left|
	\partial_k \Delta_{mn}(k)
	\right|_{k=k_i}
	},
\end{align}
with
\begin{align}
	\mathcal W_{nml}(k_i)
	= - \gamma_l
	\mathcal F_{nm}\,
	\mathrm{Im}\!\left[
		\mathcal M_{\mu\nu\lambda}^{nml}(k_i)
		\right].
\end{align}

On the other hand, the 1D joint density of states associated with the transition energy $\Delta_{mn}(k)$ is
\begin{align}
	D_{mn}^{\mathrm{1D}}(\hbar\omega)
	 & =
	\frac{1}{2\pi}
	\int dk\,
	\delta\!\left[
		\hbar\omega-\Delta_{mn}(k)
	\right]  \nonumber \\
	 & =
	\frac{1}{2\pi}
	\sum_{k_i}
	\frac{1}
	{
	\left|
	\partial_k \Delta_{mn}(k)
	\right|_{k=k_i}
	}.
	\label{eq:appendix_DOS}
\end{align}
Comparing Eqs.~\eqref{eq:appendix_conductivity} and \eqref{eq:appendix_DOS}, we find that the resonant contribution to the spin conductivity has the same singular structure as the one-dimensional joint density of states, while it is weighted by the matrix-element factor $\mathcal W_{nml}(k_i)$. Therefore, when $\mathcal W_{nml}(k_i)$ varies slowly near the resonant momenta, the conductivity approximately follows the behavior of the density of states. In contrast, the conductivity cannot follow the DOS precisely in general.

In addition, the conductivity in the 3D case can be simplified as
\begin{align}
	\mathrm{Re}\left[\sigma_{\mu\nu\lambda}^{(2)}\right]
	\propto
	\sum_{nml}
	\int_{\Delta_{nm}(\bm k)=\hbar\omega}
	\frac{
		dS_{\bm k}\mathcal W_{nml}(\bm k)
	}
	{
		|\nabla_{\bm k}\Delta_{mn}(\bm k)|
	},
\end{align}
and the joint DOS reads
\begin{align}
	D_{mn}^{\mathrm{3D}}(\hbar\omega) =
	\frac{1}{(2\pi)^3}
	\int_{\Delta_{mn}(\mathbf{k})=\hbar\omega}
	\frac{dS_{\mathbf{k}}}
	{\left|\nabla_{\mathbf{k}}\Delta_{mn}(\mathbf{k})\right|}.
\end{align}
Although the DOS in three dimensions can be a complicated function, the conductivity still has a close relation to the DOS.

%
%
%
\bibliography{ref} 

\end{document}